\documentclass[showpacs,preprintnumbers,amsmath,amssymb]{revtex4}
\def\bk{{\bf k}}
\def\bq{{\bf q}}

\usepackage{graphicx}
\usepackage{dcolumn}
\usepackage{bm}

\begin{document}

\title{Electron-phonon interaction in the solid form of the smallest fullerene C$_{20}$ }

\author{I. Spagnolatti, M. Bernasconi, and G. Benedek}
\affiliation{Dipartimento di Scienza dei Materiali and Istituto 
Nazionale di Fisica per la Materia -  Universit\`a di  Milano-Bicocca, 
Via Cozzi 53, I-20125, Milano, Italy}

\begin{abstract}
The electron-phonon coupling of a theoretically devised carbon phase made by assembling the
smallest fullerenes C$_{20}$ is calculated from first principles.
The structure consists of C$_{20}$ cages in an {\it fcc} lattice interlinked by two bridging carbon atoms
in the interstitial tetrahedral sites ({\it fcc}-C$_{22}$). The crystal is insulating but can be made metallic by 
doping with interstitial alkali atoms. 
In the compound NaC$_{22}$
the calculated  coupling constant $\lambda/N(0)$ is 0.28 eV, a value
much larger than in C$_{60}$, as expected from the larger curvature of C$_{20}$. 
On the basis of the McMillan's formula, the calculated $\lambda$=1.12 and a $\mu^*$ assumed in the range 0.3-0.1
a superconducting T$_c$ in the range 15-55 K is predicted.
\end{abstract}


\maketitle

\begin{figure}
\vskip -8 truecm
\includegraphics{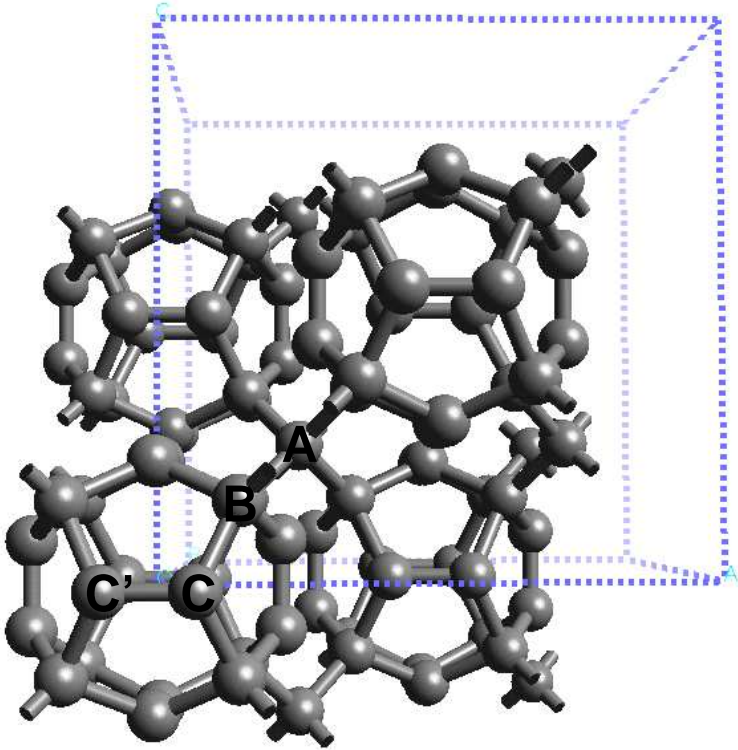}
\caption{The face-centered-cubic structure of the
{\it fcc}-C$_{22}$ crystal. The conventional unit cell with four formula 
units is shown. 
The crystal belongs to the Fm\=3 space group and the coordinates of the
three independent atoms in units of the lattice constant are 
$A \equiv$ (0.25, 0.25, 0.25), $B \equiv$ (0.148, -0.148, 0.148), 
$C \equiv$ (0.078, -0.217, 0.0). 
The atoms $C$ and $C^\prime$ and the other 10
of the cage equivalent by symmetry are linked by an ethylenic-like 
bond 1.346 $\rm\AA$ long. 
In the NaC$_{22}$ compound the alkali atom occupy the 
octahedral site of the {\it fcc} lattice.}
\label{c20fig}
\end{figure}

The discovery of high superconducting transition temperature in alkali-doped C$_{60}$ ($T_c$=10-40 K) has attracted
much attention on these materials as representatives of  a new class of superconductors \cite{C60,gunnarsson}.
Superconductivity, albeit at lower temperature ($T_c$=7 K), has been recently observed in another member of this
class, C$_{70}$, which indicates the possibility that other fullerenes might also
become superconducting \cite{batlogg_C70}. 
In this respect, the smaller fullerenes are the most interesting since the
electron-phonon coupling (and hopefully $T_c$) has been calculated to  become larger with increasing curvature
  of the molecule \cite{tomanek,cote2,lannoo,colo}. Therefore, the recent synthesis of the smallest fullerene, having the largest curvature, C$_{20}$, starting from the stable C$_{20}$H$_{20}$ molecules \cite{prinzbach}
represents a step further along the way towards high-$T_c$ superconductivity in carbon-based materials.
Experimental evidence for the realization of the next step, the
 formation of a crystalline solid based on C$_{20}$ clusters, has been recently
 provided  as well \cite{Iqbal}. Crystalline micro-domains of the new 
carbon phase have been detected by
Trasmission Electron Microscopy (TEM) in carbon films deposited on a nickel substrate by the ultraviolet laser
ablation of diamond. A face-centered-cubic  ({\it fcc}) symmetry  have
been inferred from TEM data. The Raman spectrum  shows a strong peak at 1618 cm$^{-1}$ typical
of the stretching of the ethylenic C=C double bond and the presence of C$_{20}$ clusters as building blocks of
the new crystal has been determined by mass spectroscopy of laser-desorbed fragments \cite{Iqbal}.
Based on this information we have theoretically conceived a C$_{20}$-based solid whose calculated
structural and vibrational properties are suitable to reproduce the experimental TEM and Raman data \cite{nostro}.
The structure comprises C$_{20}$ cages in a {\it fcc} lattice interconnected by two bridging atoms per unit cell in the
tetrahedral interstitial sites (Fig. \ref{c20fig}). Hereafter it  is referred to  as {\it fcc}-C$_{22}$, 
the unit cell containing 22 atoms. 

In this letter, we report on the ab-initio study of the electronic properties of {\it fcc}-C$_{22}$ doped with 
alkali metals (Na, Li)
in the search for metallization and superconductivity. The calculated electron-phonon coupling constant $\lambda$ for
the alkali-doped {\it fcc}-C$_{22}$ is found to be substantially enhanced compared to C$_{60}$ leading to the
possibility of larger $T_c$ than in alkali-doped fullerite.

We use density functional theory in the local density approximation (LDA), norm-conserving pseudopotentials and
plane-wave expansion of the Kohn-Sham orbitals up to a kinetic cutoff of 40 Ry \cite{notaLDA}. Non-linear-core
corrections have been added for the alkali ions.  Phonon frequencies and electron-phonon coupling constants have
been computed within the linear response theory \cite{RMP_baroni,Pwscf}. 
This approach has been successfully applied to the calculation of phonon dispersion relations in insulator and metals
\cite{RMP_baroni,degironc} and of the electron-phonon coupling constant $\lambda$ in simple 
and transition metals \cite{pavone}.
Brillouin Zone (BZ) integrations are performed using Hermite-Gaussian smearing of order one \cite{degironc}
with a linewidth of 0.01 Ry and 18 special points in the irreducible wedge of the BZ (IBZ).
  
The calculation of the equation of state of {\it fcc}-C$_{22}$ 
yields a cohesive energy of 0.63 eV/atom lower than that of diamond and a
lattice constant of 8.61 $\rm\AA$.
The density is 1.78 g/cm$^3$, close to that of fullerite (1.68 g/cm$^3$), but the bulk modulus is as large as 239
 GPa due to the presence of covalent interlinks between adjacent C$_{20}$ cages.
The crystal belongs to the Fm\=3 space group and contains only three independent atoms in the unit cells, the
other being obtained by symmetry operations. Their coordinates 
 in units of the lattice constant are 
 $A \equiv$ (0.25, 0.25, 0.25), $B \equiv$ (0.148, -0.148, 0.148), $C \equiv$ (0.078, -0.217, 0.0) \cite{nostro,notasaito} (Fig. 1).
Eight atoms out of twenty of the cage (atoms $B$) and the interstitial carbon atoms ($A$) are sp$^3$ hydridized.
The other twelve atoms of the cage ($C$) are sp$^2$ hybridized forming C=C ethylenic bonds. 
There are three bond lengths in the crystal:
the single sp$^3$-like $B-C$ bond in the pentagonal ring (1.531 $\rm\AA$), the C=C ethylenic-like bond
(1.346 $\rm\AA$)
long and the single sp$^3$-like bond between a cage (equivalent to $B$) and an interstitial atom ($A$) (1.517 $\rm\AA$).

\begin{figure}
\vskip -8 truecm
\includegraphics{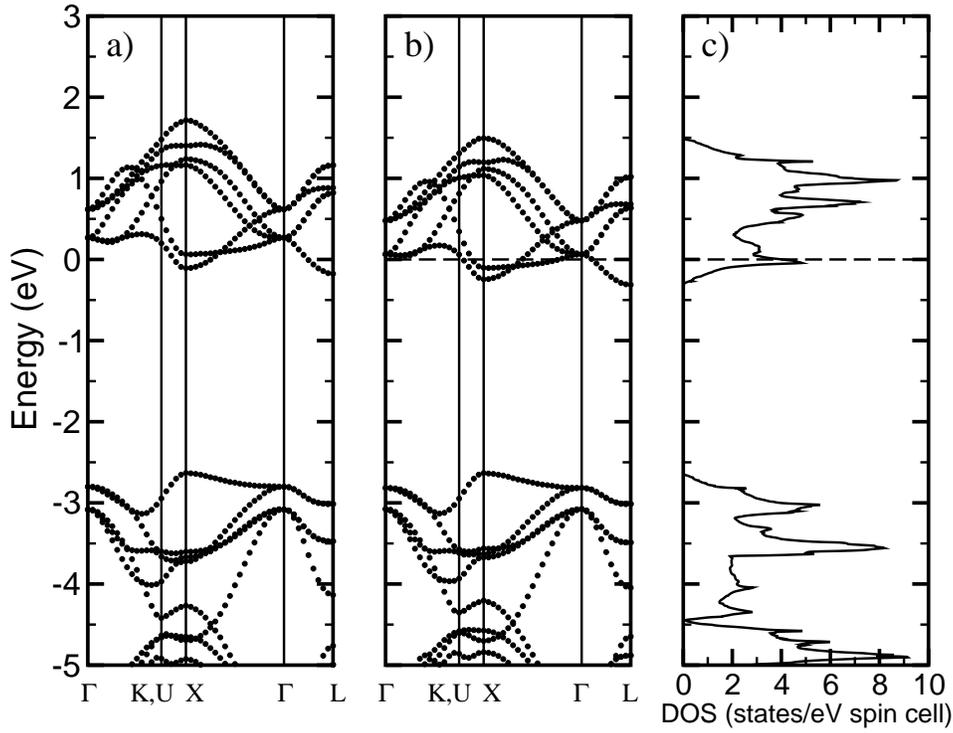}
\caption{Band structure around the Fermi level  of the pure 
{\it fcc}-C$_{22}$ (a) and NaC$_{22}$ (b) crystals.
The zero of energy is the Fermi level of NaC$_{22}$. The top of the valence bands
of the two compounds is aligned.
The two lowest
conduction bands at the $\Gamma$ points have t$_u$ simmetry. 
The electronic density of state of NaC$_{22}$ is also reported (c), as 
computed by the  tetrahedron method with 213 points in the IBZ.}
\label{bande22}
\end{figure}

The electronic band structure of {\it fcc}-C$_{22}$ is reported in Fig.
\ref{bande22}. The system is an insulator with 
a slightly indirect $XL$ band gap of 2.47 eV.
Projection of the Kohn-Sham states on the minimal atomic basis set shows that
the lowest conduction bands and the highest valence bands are mainly formed by $\pi^*$ and $\pi$ states of the 
ethylenic-like bonds.

It is suggestive to wonder if doping might turn {\it fcc}-C$_{22}$ metallic by electron transfer into the molecular-like lowest conduction
bands, as occurs for instance in fullerite. The analogy with C$_{60}$ is evident by considering that both 
C$_{60}$ molecules in
fullerite and C$_{20}$ cages in {\it fcc}-C$_{22}$ form a {\it fcc} lattice. In the A$_3$C$_{60}$ (A=K, Cs, Rb) compounds 
the three alkali metal ions per unit cell occupy the two tetrahedral sites and the octahedral site \cite{gunnarsson}.
In {\it fcc}-C$_{22}$ the two tetrahedral sites are occupied by the interstitial carbon atoms, but the octahedral
 site is empty and might host a guest ion. 
In the search for metallization we have inserted a Na atom per unit cell in the octahedral site.
Calculation of the equation of state of the compound NaC$_{22}$
shows an increase in the lattice parameter  ($a$=8.765 $\rm\AA$)
 with respect to the pure phase
and a larger (2 $\%$) increase in the C=C bond length due to population of antibonding $\pi^*$ states. 
In fact,  Na is ionized by
 donating its outermost electron to the host which becomes metallic. The Na-host interaction is weak and 
the band structure of
NaC$_{22}$ is very close to that of the pure system, but for the shift of the Fermi level inside the lowest
conduction bands (Fig. \ref{bande22}b). 
Therefore the rigid band approximation holds to a large extent in the description of
Na  insertion in {\it fcc}-C$_{22}$. 
 Although the lowest conduction band is 3-fold degenerate at the $\Gamma$ point (Fig. \ref{bande22}a),
the Fm\=3 symmetry is preserved in NaC$_{22}$ and
no static Jahn-Teller distorsion is observed. This is due to the fact that the minima of the
lowest conduction bands are located at the $X$ and $L$ points where the degeneracy is already lifted. 

To study the superconducting properties we have computed the electron phonon coupling constant $\lambda$ as

\begin{eqnarray}
  \frac{\lambda}{N(0)} & = & \sum_{\alpha} \int \frac{d \bq}{\Omega_{\rm{BZ}}}
 \frac{\lambda_{\alpha\bq}}{N(0)}   \\
& = & \sum_{\alpha} \int \frac{d \bq}{\Omega_{\rm{BZ}}} \frac{1}
 {\omega^2_{\alpha\bq}} \sum_{n,m}
\int \frac{d \bk}{\Omega_{BZ}} 
\frac{\delta (E_{\bk n}) \delta (E_{\bk + \bq m} )}{N^2(0)}  \mid
\langle u_{\bk + \bq,m}  \mid {\bf M}^{-\frac{1}{2}} {\bf \epsilon}_{\alpha \bq} 
 {\bf \nabla} V_{eff}^{\bq} \mid u_{\bk,n} \rangle \mid^2 \nonumber
\label{lambdaBZ}
\end{eqnarray}

\noindent where $N(0)$ is the electronic density of states at the Fermi level (zero of energy) 
per spin and per cell,
$\lambda_{\alpha\bq}/N(0)$ is the partial electron-phonon coupling due the $\alpha$-th
phonon at the point $\bq$ in the BZ (of volume $\Omega_{\rm{BZ}}$),
 ${\bf \epsilon}_{\alpha \bq}$ is the phonon polarization vector,
${\bf M}$ is the atomic mass matrix, $n$ and $m$ are electronic band indices, $u_{\bk,n}$ is the periodic part
of the Kohn-Sham state with energy $E_{\bk n}$, and ${\bf \nabla} V_{eff}^{\bq}$ is the derivative of the
Kohn-Sham effective potential with respect to the atomic displacement caused by a phonon with 
wavevector $\bq$. We replaced the $\delta$-functions in Eq. \ref{lambdaBZ} with Hermite-Gaussian forms 
of 0.01 Ry linewidth.
The BZ integration over $\bk$ is performed on a grid of 18 special points in the IBZ.
We computed $\lambda_{\alpha\bq}/N(0)$ for $\bq$ along the high symmetry directions [100]
and [111]. The integration over $\bq$ in Eq. \ref{lambdaBZ} is performed over a uniform mesh in
the IBZ by assigning to each point the value of $\lambda_{\alpha\bq}/N(0)$ computed for the closest
point along a high symmetry directions \cite{nota_converg}. 
Finally, for NaC$_{22}$
we obtain $\lambda/N(0)$=0.28 eV, $N(0)$=4.0 states/eV/spin/cell (as computed by the tetrahedron 
method with 213 points in the IBZ)
 and $\lambda$=1.12. A similar calculation for the charged fullerite C$^{3-}_{60}$
at the $\Gamma$ point only in the molecular-like approximation \cite{nota_molecular} gives
$\lambda/N(0)$=0.076 eV, in agreement with previous works \cite{gunnarsson}
(or 0.064 eV by including the set of phonons considered in table IV of Gunnarsson's review
\cite{gunnarsson}). 
Therefore, although {\it fcc}-C$_{22}$ is not strictly a fullerite, the C$_{20}$ clusters
 being covalently interlinked
via  interstitial atoms, the solid retains some features of its
 molecular building blocks, giving rise to
a substantial enhancement of $\lambda/N(0)$ with respect to C$_{60}$ as expected from the 
larger curvature of C$_{20}$ \cite{lannoo}.  
The persistence of molecular features in the crystal is also confirmed by the calculation of
 $\lambda/N(0)$ in the molecular-like approximation \cite{nota_molecular} for the
pure {\it fcc}-C$_{22}$ doped with an additional electron (equally shared by the three lowest
$t_u$ states). This latter calculation gives $\lambda/N(0)$=0.31 eV, a value very close to the 
result  for NaC$_{22}$ at convergence in BZ integration (0.28 eV) \cite{nota_litio}.
A tight-binding calculation of the electron-phonon interaction in the isolated C$_{20}$ cluster gave 
a somehow lower value $\lambda/N(0)=0.18$ eV \cite{lannoo} which may be partially due to the semiempirical method
used.

\begin{figure}
\vskip -20 truecm
\includegraphics{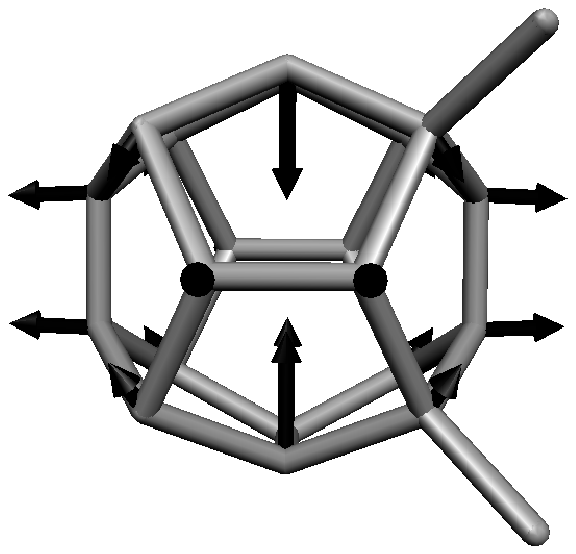}
\caption{Displacement pattern of the $E_g$(1) phonon at 334 cm$^{-1}$ (cfr. Table I).}
\label{Egphon}
\end{figure}

In order to identify the phonons which mostly contribute to the elctron-phonon coupling, we 
report in
 Table I the partial electron-phonon
coupling constants $\lambda_{\alpha}/N(0)$ of
 the individual $\alpha$-th phononic band, obtained by averaging $\lambda_{\alpha \bq}/N(0)$  over the BZ.
The frequency and symmetry of the $\alpha$-th phononic bands at the $\Gamma$ point 
are also given in Table I for  the pure {\it fcc}-C$_{22}$ and NaC$_{22}$ phases.
Although the phonons with the largest matrix elements in Eq. \ref{lambdaBZ} correspond
to the stretching modes of the C=C ethylenic bonds (1359-1362 cm$^{-1}$), their contribution to
$\lambda$ is attenuated by the factor $1/\omega^2$. As a consequence the largest  $\lambda_{\alpha}/N(0)$
is due to the $E_g$ mode  at 334 cm$^{-1}$ (cfr. Table I);
its displacement pattern (Fig. \ref{Egphon}) produces
the largest modulation of the dihedral angles which control the $\pi$ conjugation of
the ethylenic bonds. This is at variance with C$_{60}$ where the largest electron-phonon coupling constants
are due to the phonons with the highest frequencies \cite{gunnarsson}.

In system where the superconductivity is driven by the electron-phonon 
interaction an estimate of $T_c$ can be obtained by making use of the 
McMillan's solution \cite{carbotte,allen} of
the Eliashberg equations:

\begin{equation}
T_c= \frac{\omega_{ln}}{1.2} {\rm exp} \biggl( - \frac{1.04(1+\lambda)}
{\lambda - \mu^* (1+0.62 \lambda)} \biggr)
\label{Tc}
\end{equation}

where $\omega_{ln}$ is the weighted logarithmic average of the phonon frequencies \cite{carbotte,allen} (470 cm$^{-1}$ in NaC$_{22}$)
and $\mu^*$ is the Coulomb pseudopotential $\mu=U N(0)$ renormalized by retardation effects.
$U$ is a typical screened Coulomb interaction among states at the Fermi level \cite{allen}.
Latest estimates give $\mu^* \sim$ 0.3 for C$_{60}$ although, due to the 
similarity between the electronic bandwidth (0.5 eV) and phonon energies 
(up to 0.2 eV), the applicability 
of Eq. \ref{Tc} is questionable for fullerite \cite{gunnarsson} .
We have not attempted to estimate numerically $\mu^*$ for NaC$_{22}$, so our discussion now turns on a very speculative
level. Although $U$ might be larger in NaC$_{22}$ than in fullerite due to the smaller size of
C$_{20}$, $N(0)$ turns out to be lower
(4.0 and 6.6-9.8 states/eV/spin/cell in NaC$_{22}$ and K$_3$C$_{60}$ from band structure 
calculations \cite{gunnarsson}, respectively)
due to the larger dispersion of the conduction bands, three times larger than in C$_{60}$.
The estimate of the renormalization of $\mu$ leading to $\mu^*$ is difficult
\cite{gunnarsson}, however the renormalization is perhaps larger than in C$_{60}$ due to the larger bandwidth.
Thus, if we assume that the value
of  0.3 estimated for fullerite is an upper bound for $\mu^*$ in NaC$_{22}$ 
we obtain $T_c$=15K
from Eq. \ref{Tc} and $\lambda$=1.12. For the more optimistic case of
$\mu^*$ in the range 0.1-0.2 typical of simple and transition metals \cite{carbotte,allen}, we obtain 
the remarkable result of $T_c$=55-33 K ($T_c$ changes linearly with $\mu^*$ in this range). 

Since $\lambda/N(0)$ is mainly a molecular property, we have 
also attempted to increase $\lambda$ by
increasing $N(0)$ with a lattice expansion. This has been achieved by substituting C by Si in the interstitial sites
producing the compound NaSi$_2$C$_{20}$. This latter system has a substantially expanded lattice parameter
($a$=9.49 $\rm\AA$, $N(0)$=4.44 states/eV/spin/cell) 
but only a marginally enhanced $\lambda$=1.49 with respect to NaC$_{22}$. 
In fact the increase in $N(0)$ is lower than
expected from the lattice expansion due to the reduction of bands at the Fermi level from 
two in NaC$_{22}$ (cfr. Fig. \ref{bande22}) to one in NaSi$_2$C$_{20}$. Moreover, a
decrease in $\omega_{ln}$ due to the lattice expansion compensates the increase in $\lambda$ and finally  yields a
$T_c$ close (within 10 K) to that of NaC$_{20}$.
However, further increase in $\lambda$ is
foreseeable for the compounds NaGe$_2$C$_{20}$ and NaSn$_2$C$_{20}$.

In summary, we have carried out ab-initio calculations on the electronic and  superconductive properties
of a new carbon phase {\it fcc}-C$_{22}$ formed by assembling C$_{20}$ clusters in a {\it fcc} lattice.
The system is insulating but can be made metallic by doping with interstitial alkali metals.
Although the structure is not a molecular crystal, the C$_{20}$ cages being covalently interlinked via interstitial
carbon atoms, the crystal  retains some molecular features. In fact, as expected from the molecular properties of the isolated 
C$_{20}$ cluster, the compound NaC$_{22}$ exhibits a large electron-phonon coupling constant $\lambda/N(0)=0.28$ eV, 
strongly enhanced with respect to C$_{60}$ due to the larger curvature of C$_{20}$. Estimate of $N(0)$ from band structure
calculation yields $\lambda=$1.12 which allows a rough estimate of $T_c$ from McMillan's formula.
For $\mu^*$ in the reasonable range of 0.3-0.1 we predict $T_c$=15-55 K for NaC$_{22}$.
Although a more precise estimate of $T_c$ requires the difficult calculation of $\mu^*$, the results presented here 
shows that NaC$_{22}$ is a potentially interesting superconductor.
We hope that these theoretical results could stimulate further experimental work on the synthesis and
characterization of this new structure.

This work is partially supported by the INFM Parallel Computing Initiative,
and by MURST, through COFIN99 and COFIN01 under contract  2001021133.
We gratefully aknowledge Z. Iqbal for discussion, information and for providing us with
his experimental data prior to publication.

\begin{table}
\caption{The partial electron-phonon coupling constant $\lambda_{\alpha}/N(0)$ (eV)
of the individual $\alpha$-th phononic band, obtained by averaging $\lambda_{\alpha \bq}/N(0)$  over the BZ.
 The frequency and simmetry of the $\alpha$-th phononic bands at the $\Gamma$ point
  are also given for the pure {\it fcc}-C$_{22}$ and NaC$_{22}$
 phases. Only the phononic bands with
   $\lambda_{\alpha}/N(0)$ larger than 5 meV are reported.
 T$_u$(1) correspond to the acoustic bands. T$_u$(2) correspond
 to translational modes of Na ions.}
\begin{center}
\begin{tabular}{lcc}
Modes  & Energy ($cm^{-1}$)  & $\lambda_{\alpha}/N(0)$ \\
& C$_{22}$ $\; \;$     NaC$_{22}$ & \\
\hline
T$_u$(1) &   -  $\; \;$    -  & 0.010 \\
T$_u$(2) &   -  $\; \;$   271 & 0.022 \\
T$_g$(1) &  281 $\; \;$   282 & 0.046 \\
T$_g$(2) &  490 $\; \;$   445 & 0.038 \\
E$_g$(1) &  492 $\; \;$   334 & 0.072 \\
T$_u$(3) &  576 $\; \;$   482 & 0.019 \\
T$_u$(4) &  626 $\; \;$   609 & 0.008 \\
T$_g$(3) &  701 $\; \;$   637 & 0.008 \\
T$_u$(5) &  731 $\; \;$   697 & 0.007 \\
E$_u$(1) &  754 $\; \;$   719 & 0.005 \\
T$_g$(4) &  857 $\; \;$   823 & 0.005 \\
T$_u$(6) &  948 $\; \;$   927 & 0.006 \\
T$_u$(10)& 1559 $\; \;$  1362 & 0.010 \\
E$_g$(3) & 1574 $\; \;$  1359 & 0.005 \\
\end{tabular}
\end{center}
\end{table}

\begin{thebibliography}{169}
\bibitem{nota_converg} To check the convergence of our results, we have considered
that $\lambda/N(0)$ represents a double average over the BZ of the matrix elements in
Eq. \ref{lambdaBZ} weighted by the function 
$f_{nm}(\bq,\bk)= \delta (E_{\bk n}) \delta (E_{\bk + \bq m})/N^2(0)$, which is
 normalized to one at convergence in the BZ integration. By normalizing the double integral in 
Eq. \ref{lambdaBZ} by the factor 
$\sum_{n,m} \int d \bq d \bk  f_{nm}(\bq,\bk)/N^2(0)$, we achieve a 
faster convergence with respect to $\bq$ and $\bk$ sums and gaussian spreading. 
We estimate a total error in $\lambda/N(0)$ below 10 $\%$.
\end{thebibliography}
\end{document}